\documentstyle[preprint,aps]{revtex}
\newcommand{\be}{\begin{equation}}
\newcommand{\ee}{\end{equation}}
\newcommand{\bea}{\begin{eqnarray}}
\newcommand{\eea}{\end{eqnarray}}
\newcommand{\bit}{\begin{itemize}}
\newcommand{\eit}{\end{itemize}}
\newcommand{\ben}{\begin{enumerate}}
\newcommand{\een}{\end{enumerate}}

\begin{document}
\title{Theory of One-Channel vs. Multi-Channel Kondo Effects for
Ce$^{3+}$ Impurities.}

\author{Tae-Suk Kim and D. L. Cox}
\address{
Department of Physics, Ohio State University, Columbus, OH 43210
}

\date{\today}
\maketitle

\widetext

\begin{abstract}
\smallskip
 We introduce a model for Ce$^{3+}$ impurities in cubic
metals which exhibits competition between the Fermi-liquid fixed
point of the single channel Kondo model and the non-Fermi-liquid
fixed point of the two- and three-channel Kondo models.
Using the non-crossing approximation and scaling theory, we find:
(i) A possible three-channel Kondo effect between
the one- and two-channel regimes in parameter space.
(ii) The sign of the thermopower is a fixed point diagnostic.
(iii) Our results will likely survive the introduction of additional
$f^2$ and conduction states. We apply this model to interpret
the non-Fermi liquid alloy La$_{1-x}$Ce$_x$Cu$_{2.2}$Si$_2$.
\smallskip
\end{abstract}
\bigskip


\pagebreak

The common paradigm to understand the electronic properties of
metals is the Landau Fermi-liquid picture, in which the low
energy excitation spectrum has a 1:1 map to that of the
non-interacting electron gas and concommitant constant specific
heat coefficient $C_{el}/T$ and susceptibility $\chi(T)$
at low temperatures.
Recent discoveries in heavy electron materials of non-Fermi
liquid behavior in $C_{el}/T$ and
other properties\cite{yupd,thursi,aliev,andraka2,unial,updal}
together with non-Fermi liquid resistance curves in small
quenched copper point contacts\cite{ralphbuh}
have stimulated renewed interest in the
multichannel Kondo model
introduced by Nozi\`{e}res and Blandin\cite{nozbland}.
In this model an impurity with internal degrees of freedom (spin $S_I$)
couples antiferromagnetically
to $M$ degenerate conduction channels or bands. For
$M> 2S_I$ the local spin is ``overcompensated''
yielding a critical ground
state with a non-Fermi liquid excitation spectrum (e.g., for
$M=2,S_I=1/2$, $C_{el}/T \sim -\ln(T)$ for $T\to
0$\cite{sacramento,conformal}).
{\it Non-magnetic} impurities with
internal degrees of freedom generically give
rise to the $M=2,S_I=1/2$ model\cite{zow,cox1}, and these models
have been argued to describe uranium based heavy fermion
alloy systems\cite{yupd,thursi,aliev,unial,updal}
and point contact data\cite{ralphbuh}. Surprisingly, to
date no realistic $M=2,S_I=1/2$ {\it magnetic} impurity
candidates have been put forward.

In this paper, we propose and study in detail an
Anderson impurity model for a Ce$^{3+}$ ion in a cubic metal
host which admits low temperature physics of either
the $M=1,2,$ or 3, $S_I=1/2$ {\it magnetic} Kondo impurity model.
The model retains the lowest crystalline electric field (CEF) states
of the $f^1$ and $f^2$ configurations.  In addition to employing perturbative
scaling arguments,
we utilize the non-crossing approximation (NCA), which we calibrate by
comparing our calculated magnetic susceptibility $\chi(T)$ curves for
$M=2,3$ with the exact results\cite{sacramento}.  We show that
a large, negative thermoelectric power $S(T)$ is a
necessary condition for observiong
the $M=2,S_I=1/2$ magnetic Kondo effect.
Our results will likely survive a
more realistic treatment including all
$f^1,f^2$ and conduction electron states.
We compare our results to La$_{1-x}$Ce$_x$Cu$_{2.2}$Si$_2$\cite{andraka2}
where $C(T)/T=\gamma(T)$
$\sim -\ln(T)$   and $\chi(T)\sim -\ln(T)$ (per
Ce ion) are observed along with large, negative $S(T)$.

The original $M=2,S_I=1/2$ Kondo model described
an impurity magnetic moment with orbital conduction channel
labels\cite{nozbland}.  Heavy electron alloys
containing trivalent cerium
(Ce$^{3+}$) ions with a lowest lying $f^1$ magnetic configuration with
a doublet CEF induced ground state
are candidate $M=2,S_I=1/2$ systems, provided two necessary conditions
are met\cite{cox1}:\\
(1) The point group of the Ce$^{3+}$ ion must be cubic or hexagonal. \\
(2) The ever-present $M=1,S_I=1/2$ coupling (induced by
$f^0-f^1$ virtual charge fluctuations)
must be smaller than the $M=2$ coupling (induced by $f^1-f^2$ virtual
charge fluctuations).
Equivalently, the ground state weight of $f^2$ must exceed that of
$f^0$.
Conditions (1,2) strongly restrict observability of the
magnetic $M=2,S_I=1/2$ non-Fermi liquid fixed point. Hence,
it is not surprising that many Ce$^{3+}$-based
heavy fermion materials are well described
by the Fermi liquid fixed point of the
$M=1,S_I=1/2$ Kondo model.

Our model Anderson Hamiltonian for Ce impurities
at a cubic site is developed by first projecting to the lowest Hund's
rule LSJ  multiplets for three configurations:
$f^0(J=0)$ with energy $\epsilon_0=0$, $f^1(J=5/2)$ with energy
$\epsilon_1$, and $f^2(J=4)$ with energy $\epsilon_2$.
By taking $\epsilon_0=0$, we have $\epsilon_2=2\epsilon_1 +U_{ff}$,
where $U_{ff}$ is the on-site Coulomb repulsion.
Further, we restrict our initial discussion to the
most strongly coupled orbital angular momentum $l_c=3$, spin $s_c=1/2$,
total angular momentum $j_c=5/2$ conduction electron partial wave
states. The conduction band density of states $N(\epsilon)$ is
taken as lorentzian
with width $D$.

With the CEF turned on, the $f^0$ configuration with only
a trivial singlet level is unaffected.
We assume the CEF induces a magnetic doublet ($\Gamma_7$)
ground state in the $f^1 J=5/2$ multiplet which
acts like an effective magnetic spin $S_I$=1/2 with indices
$\alpha=\uparrow,\downarrow$.
 In the $f^2 J=4$ multiplet, we assume a lowest
lying non-magnetic doublet ($\Gamma_3$) which has quadrupolar or shape
degrees of freedom ($n$=``+'' = stretched ion, $n$=``-''
=squashed ion). This roughly
corresponds to the $3z^2-r^2(+),x^2-y^2(-)$ orbital $E$ doublet
for transition metal ions in cubic
symmetry.  The conduction $j_c=5/2$ partial wave sextet splits
into a doublet ($\Gamma_{7c}$) and quartet ($\Gamma_{8c}$) under
the CEF.
The trivial $f^0$ symmetry implies that only $\Gamma_{7c}$
doublet partial waves induce $f^0-f^1$ mixing.
 $f^1$-$f^2$ mixing is
induced only by the $\Gamma_{8c}=\Gamma_{7}\otimes\Gamma_3$
partial waves.
This quartet is a tensor product of ``spin'' ($\Gamma_7$)
and ``orbital''
($\Gamma_3 \sim (x^2-y^2,3z^2-r^2)$) degrees of freedom.   There are
then {\it three} symmetry distinct ``channels'' of ``spin states''
($\Gamma_7$) from the conduction sector:
one from the $\Gamma_{7c}$ doublet, and two from the $\Gamma_{8c}$
quartet.

Conduction impurity coupling is effected through the
hybridization Hamiltonian, $H_{hyb}$, given by
\bea
 H_{hyb} &=& V_{01} \sum_{\epsilon \alpha} c_{\epsilon
 \alpha}^{\dagger}
      | f^0 >< f^1 \alpha | + h.c. \nonumber\\
  && + V_{12} \sum_{\epsilon n \alpha} (-1)^{\alpha+1/2}
     c_{\epsilon n \alpha}^{\dagger}
	   |f^1 \bar{\alpha} > < f^2 n | + h.c.
\eea
where
$c_{\epsilon\alpha} (c_{\epsilon\alpha}^{\dagger})$
and $c_{\epsilon n \alpha} (c_{\epsilon n \alpha}^{\dagger})$
annihilate(create) conduction electron partial wave states  of energy
$\epsilon$ in the $\Gamma_{7c}$ and $\Gamma_{8c}$ manifolds,
respectively.  All Clebsch-Gordon
coefficients are lumped into the hybridization matrix elements $V_{01}$
and $V_{12}$.
Although $V_{01}$ and $V_{12}$ are constrained by the one-body
nature of the hybridization potential, we allow their free variation
to study the competition between $M=1,2,3$ parameter regimes.
Similarly, we will not constrain $\epsilon_{1,2}$
by the known values ($\epsilon_1\approx -2$eV, $\epsilon_2\approx
-2\epsilon_1\approx 4 eV$\cite{herbst}).  We shall
critically discuss these assumptions later.

This model maps to a Kondo (exchange) form at low energy scales, using
the Schrieffer-Wolff transformation\cite{sw}.
The $S_I=1/2$ impurity spin
from the $f^1,\Gamma_7$ is coupled to conduction states
exchange integrals
 $J_1 = -2 | V_{01} |^2 / \epsilon_1 $ for $M=1,\Gamma_{7c}$ conduction
states, and $J_2 = 2 | V_{12} |^2 /
[\epsilon_2 - \epsilon_1]$ for $M=2,\Gamma_{8c}$
conduction
partial waves. When $J_1=J_2$, an $M=3$ theory results from the
combination of $\Gamma_{7c},\Gamma_{8c}$ states.
Note that the common models with $U_{ff}\to \infty$ assumed
give {\it no access} to the $M=2,3$ fixed points!  Third order
scaling theory shows that the $M=1[2](3)$ fixed points obtain
as $J_1>J_2[J_1<J_2](J_1=J_2)$, with the crossover from the high
temperature free moment fixed point set by the Kondo scale
$T_0$\cite{kimcoxtobe}.

We have studied our model with
 the non-crossing approximation (NCA), a self-consistent diagrammatic
method--for further details see Refs. \cite{bcw,coxruck}.
In the NCA
pseudo-particle Green's functions are introduced for each ionic state
of the $f^{0,1,2}$
configurations, and the self-energy equations for these propagators are
self-consistently solved to second order in $V_{01},V_{12}$.  The NCA
yields the correct critical behavior for the
over-screened $SU(N)\otimes SU(M)$ multi-channel Kondo model, with $N$
the impurity spin degeneracy\cite{coxruck}. The NCA
produces pathological singularities for $M=1$,
but only below a scale $T_p<<T_0$\cite{bcw}. In particular,
 $\chi(0),\gamma(0) \sim 1/T_0$ as expected.
Hence, the NCA is a reliable method for
detailed studies of the competition between $M=1,2,3$ fixed points.

At $T=0$, the NCA can be reduced to coupled non-linear
differential equations for
the inverse configuration propagators as a function of frequency which
can be solved analytically close to a threshold energy
($E_0\approx \epsilon_1+O(V_{01},V_{12})^2$)\cite{bcw,coxruck}.
We introduce the quantity
$\gamma_c = \pi[(\epsilon_2 - E_0) / \Gamma_{12}
  +   E_0 / \Gamma_{01}]$$= (\pi\tilde \epsilon_2/\Gamma_{12})$
which also defines a renormalized $f^2$ energy $\tilde\epsilon_2$,
where
$\Gamma_{ij} = \pi N(0)V_{ij}^2$. Clearly, $\gamma_c\approx
(N(0)J_2)^{-1}-(N(0)J_1)^{-1}$.
For $\gamma_c\sim \tilde\epsilon_2
> 0[<0](=0)$, we obtain $M=1[M=2](M=3)$ low temperature
physics, in agreement with the third order scaling in $J_1,J_2$
mentioned above.

 Parameters for our NCA work are listed in Table 1.
For simplicity,  we take $\Gamma=\Gamma_{12}=\Gamma_{01}$, so
$\epsilon_2$ alone regulates the $T\to 0$ physics. In
addition, we can analytically
estimate $T_0$ (here $\epsilon_{12}=\epsilon_1-\epsilon_2$):\\
$\bf {\it M}=1,\epsilon_2>0$:
$k_BT_0 = \epsilon_2(\Gamma/\pi D)^{1/2}
\exp(\pi\epsilon_1/ 2\Gamma)$;\\
$\bf {\it M}=2,\epsilon_2<0$:
$k_BT_0 = (\Gamma/\pi)(|\epsilon_2|/ D)^{1/2}
\exp(\pi\epsilon_{12}/2\Gamma) $;\\
$\bf {\it M}=3,\epsilon_3=0$:
$k_BT_0 = D(\Gamma/\pi D)^{3/2} \exp(\pi\epsilon_1/2\Gamma)$.\\
In deriving  these $T_0$ expressions, we  assume
$\Gamma,|\epsilon_{1,2}|<<D$.
For $|\gamma_c|<<1$, the scaling trajectories flow
close to the $M=3$ fixed point, crossing over to
the $M=1,2$ fixed points below
a scale $T_x \sim |\epsilon_2|^{5/2}$\cite{kimcoxtobe}.

Calculated $\chi(T)$ curves allow us to test the NCA and develop
a model phase diagram as shown in Fig. 1.
$\chi(T)$ diverges as $-\ln(T/T_0)[(T/T_0)^{-1/5}]$
for the $M=2$[$M=3$]-channel
$S_I=1/2$ model as $T\to 0$\cite{sacramento,conformal},
while for $M=1$, $\chi(0)\sim 1/T_0$ at zero temperature.
Our NCA results show approximate scaling behavior
for all the $M=1,2,3$ parameter sets, and
agree excellently with exact Bethe ansatz
results for $M=2,3$\cite{sacramento} as anticipated from $T=0$
NCA analysis\cite{coxruck} (we also find good agreement with
exact results for our $M=2,3$
$C_{el}/T$ and entropy curves\cite{kimcoxtobe}).
The $M=1$ results are less satisfying, though
$\chi(0)\sim 1/T_0$ in the NCA and negative curvature
in $\chi(T)$ is evident in the figure.
Calculation of $\chi''(\omega,T)$ for $M=2,3$ leads, as
expected\cite{coxruck} to
low $T$ marginal Fermi liquid
behavior\cite{kimcoxtobe}.

With the NCA the electrical resistivity $\rho(T)$ and
 thermopower $S(T)$ may be computed
using the Kubo formula\cite{bcw} in the dilute impurity limit,
where inter-impurity correlations can be neglected.
Assuming dominant scattering in the $l=3$ partial wave
sector\cite{bcw},
the transport coefficients are determined by the integrals
$I_n = \int d\epsilon(-\partial f/ \partial \epsilon)
\tau(\epsilon) \epsilon^n$\cite{bcw},
where $\tau(\epsilon) = \int d\hat k \tau(\hat k,\epsilon)/4\pi$ is the
angular averaged
transport lifetime determined from the one-particle $T$-matrix defined
in terms of the interconfiguration excitation propagator
\cite{kimcoxtobe,bcw,coxruck}. We obtain
the
expected behavior $\rho(T)/\rho(0) \sim 1-A(T/T_0)^{1/2}(
1-A(T/T_0)^{2/5})$
power laws for $M=2(3)$\cite{conformal,coxruck}
for $T\le 0.05T_0$\cite{kimcoxtobe,coxmak}.

New physics is evident in $S(T)=-I_1/eTI_0$,
with $e$ the electron charge. Given $T_0<<D$, the sign of $S(T)$
is determined by the degree of particle-hole asymmetry in
$1/\tau(\epsilon)$.
Figure 2 displays the differing $S(T)$ behaviour for the $M=1,2,3$
regimes of
our model.  $S(T)$ is negative definite in the $M=2,3$ regimes of the
model\cite{kimcoxtobe}.
For $M=1$, dominant $f^0$-$f^1$ mixing yields a Kondo resonance
peak in $1/\tau$ just above the Fermi level, giving
$S(T)>0$.
On the other hand, $S(T)<0$ arises when scattering between the
$f^1$ and $f^2$ sectors dominates, since
the overall resonance spectral weight in $1/\tau$
is shifted below the Fermi energy, giving stronger hole than
electron scattering for $T\to 0$.
Hence, $S(T)<0$ for low temperatures
when Condition (2) for the $M=2,S_I=1/2$ magnetic Kondo
effect is satisfied.

We now address the effect of omitted $f^2$ and conduction states:\\
{\it (1) Contributions of excited $f^2,\Gamma_3$ states}
We estimate $J_2/J_1\approx 0.2$ for
realistic $\Gamma_{12},\Gamma_{01},\epsilon_2,\epsilon_1$ values and
inclusion of the lowest $f^2 \Gamma_3$ state.
However, we can make $J_2>J_1$
within a spherical hybridization model (and Lorentzian density
of states) by: (a) including all nine
$f^2\Gamma_3$ states, which effectively enhances $V_{12}^2$ by a
factor of five, and (b) reducing the conduction occupancy from
1.0 to 0.2 per site, qualitatively consistent with electronic
structure inputs to analysis of spectroscopic data for
CeCu$_2$Si$_2$(see Ref. \cite{allen}(a)). \\
{\it(2) New contributions from other excited levels}.
Virtual charge fluctuations to excited $f^2$ triplet states suppress
the one-channel coupling and
induce a new {\it channel symmetry breaking} exchange interaction
of the form
$\tilde H = \tilde J \sum_{\lambda}S_7^{\lambda}(\tau^{\lambda}
S_8^{\lambda})$
where $S_7^{\lambda},S_8^{\lambda}$ are pseudo-spin matrices for the
$\Gamma_7$
impurity and $\Gamma_8$ conduction states, and the $\tau^\lambda$
matrices from the
$\Gamma_8$ channel space have the forms
$\tau^x=(-\tau^{(3)}-\sqrt{3}\tau^{(1)})/2$,$\tau^y=
(-\tau^{(3)}+\sqrt{3}\tau^{(1)})/2$, and $\tau^z=\tau^{(3)}$ following
Pauli matrix conventions.
Third-order scaling and strong coupling analysis
show that the $M=2$ fixed point for $J_2>J_1$ is
stable provided  $2 J_2 > |\tilde J|$, which is
possible for realistic values of
$V_{01},V_{12},\epsilon_1,\epsilon_2$.\\
{\it (3) Additional partial wave  states}.
An added $j_c=7/2\Gamma_{8c}$ will mix
with the $j_c=5/2,\Gamma_{8c}$ quartet resulting in new exchange terms.
A unitary transformation
to bonding and non-bonding quartets yields two identical, decoupled
$M=2,S_I=1/2$ equations with enhanced coupling strength  in the bonding
channel. The non-bonding coupling has a zero bare value, and so is
practically irrelevant.

Turning to the alloy, La$_{1-x}$Ce$_x$Cu$_{2.2}$Si$_2$, it is found that
$\chi(T),\gamma(T)\sim -\ln(T)$ for $x=0.1$\cite{andraka2}.
We thus examine whether it meets the necessary Conditions
(1,2) for the $M=2,S_I=1/2$ fixed point.
First,
the Ce ions sit in a pseudo-cubic environment with a doublet ground
state well known from neutron scattering\cite{osborne}.
The best superconducting samples of CeCu$_{2.2}$Si$_2$ have isotropic
low temperature susceptibilities\cite{stegrewe}.  Hence, condition (1)
appears to be satisfied.
Using data from Ref. \cite{andraka2} and the cubic $\Gamma_7$ effective
moment to estimate the Landau-Wilson ratio $R$ given by
\begin{equation}
R=\left[{\pi^2 k_B^2 (\partial\chi(T)/\partial\ln T)\over
\mu_{eff}^2 (\partial\gamma(T)/\partial\ln T)}\right]_{T\to 0}
\end{equation}
we obtain $R=$2.7(1), in excellent agreement with
 the theoretical $M=2,S_I=1/2$ value
$R=8/3$\cite{sacramento,conformal}.  The uncertainty
reflects measurement
accuracy from the plots of Ref. \cite{andraka2}.
Further support arises from the enhancement of $\gamma(T)$
in applied magnetic field\cite{andraka2}
which is qualitatively (though
not quantitatively) consistent with the $M=2,S_I=1/2$
Bethe-Ansatz results\cite{sacramento}; in contrast, for
the $M=1,S_I=1/2$ model, $\gamma(T)$ drops with applied field.
The resistivity $\rho(T)$ does not display clean
$T^{1/2}$ behavior, but has been measured down to only
$T\approx T_0/10$=1.2K\cite{andraka2}; our work shows that
the $T^{1/2}$ behavior sets in below about
0.05$T_0$\cite{kimcoxtobe,coxmak}.

Turning to condition (2), we consider
the $x=1$ $S(T)$ data for which a
sign change occurs
at 70K, well above $T_0\approx 10K$, eventually reaching a large
negative value of
about $-20-30 \mu$V/K, which is in reasonable agreement with our $M=2$
curves.
We observe that: (i) For $x=0$ the thermopower is positive, positively
identifying the sign
change with the Ce ions. (ii) Coherent lattice
quasiparticle effects can only be expected for $T<<T_0\approx 10K$ (in
other Ce based
heavy fermion materials, sign changes typically occur well below $T_0$
with magnitudes of only a few $\mu$V/K).  (iii) The sign
change occurs well below the CEF excitation energy of
360K\cite{stegrewe,jaccard}, so is not due to CEF effects (although
the positive $S(T)$ values for
higher $T$ are likely due to scattering
off the excited CEF quartet, which generates a Kondo resonance above
the Fermi level). (iv) The magnitude is easily explained only with a resonant
scattering mechanism near the Fermi level.  Observations (i)-(iv)
strongly suggest that
the negative thermopower arises from the single ion Kondo
effect
with $f^1-f^2$ fluctuations dominating the virtual exchange
processes, hence satisfying
necessary condition (2), which is further supported by analysis
of high energy spectroscopies\cite{allen}.
Negative $S(T)$ are measured at $T=20K$ for many
Ce based 1-2-2 structures with high unit cell
volumes, while positive $S(T)$ at $T=20K$ are seen for low
volumes\cite{jaccard}.
Indeed, increasing pressure
can tune from $M=2$ through $M=3$ to $M=1$ fixed points
since pressure initially destabilizes $f^2$
relative to $f^0$\cite{herbst}.

In summary, we believe the
La$_{1-x}$Ce$_x$Cu$_{2.2}$Si$_2$
system is a promising $M=2,S_I=1/2$ (or $M=3$ under pressure)
Kondo candidate which should be studied further, particularly
because magnetic Kondo impurities are more readily studied
(they couple directly
to magnetic field and neutrons).

This research was supported by a grant
from the U.S. Department of Energy, Office of Basic Energy Sciences,
Division of Materials
Research.  We thank R. Heid, L.N. Oliveira,
and J.W. Wilkins for stimulating
interactions.

\begin{table}
\begin{tabular}{|c||c|c|c|c|c|} \hline
Set  & $M$ & $\Gamma/D$  &  $\epsilon_1/D$ &
 $\epsilon_2/D$ & $k_BT_0/D$ \\ \hline \hline
 1 & 2 &0.2  & -0.4 & -0.1 & .19081E-02 \\
 2 & 2 & 0.2 & -0.37 & -0.07 & .15964E-02 \\
 3 & 2 &0.2  &  -0.35 & -0.05 & .13492E-02 \\
\hline
 4 & 3 & 0.2 & -0.3 & 0.0 & .15224E-02 \\
 5 & 3 & 0.2 & -0.4 & 0.0 & .69413E-03\\
\hline
 6 & 1  & 0.2 & -0.3 & 0.05 & 0.11957E-02 \\
 7 & 1 & 0.2 & -0.3 & 0.07 & 0.16740E-02 \\
 8 & 1  & 0.2 & -0.3 & 0.10 & 0.23914E-02 \\
\hline
\end{tabular}
\caption{Model parameters for the Ce impurity.
This set of model parameters covers the single, three, and
two channel Kondo regimes. $\Gamma=\Gamma_{01}=\Gamma_{12}$ is
the hybridization strength for both $f^0 - f^1$ and $f^1 - f^2$ mixing,
respectively.}
\end{table}

{\bf Figure 1. Scaling behavior for $\chi (T)$ of model Ce$^{3+}$
impurity.}
Results are shown for parameter regimes of the model (Eq. (1))
giving
$M=1,2,3$ fixed points (see Table 1).
Points: NCA calculations; Solid Lines: Exact Bethe-Ansatz
results\cite{sacramento} (with
$T_0=T_K/0.3$ for $M=2$,
$T_K$ from Ref. \cite{sacramento}).
For convenience, we multiply $\chi(T)$
by 2.0 for $M=2$.
The ground state phase diagram for the model in
exchange coupling constant
parameter space is drawn in the inset, where
$g_i =  N(0)J_i$, $N(0)$ being the conduction band density of
states at the Fermi energy. The solid diagonal
line is for $M=3$.

{\bf Figure 2. Thermopower $S(T)$.} The thermopower is
positive at low $T$ for $M=1$,
strongly
negative for $M=2$, and weakly negative for $M=3$.
Dominant $f^0-f^1$($f^1-f^2$)  virtual
charge fluctuations give positive(negative) $S(T)$.
Referring to the Table 1: the dash-dotted line is calculated with
parameter set
8, the dashed-dot-dotted line is calculated with
parameter set 1, and the solid line is calculated with
parameter set 4.

\end{document}